\documentstyle[aps,epsfig]{revtex}

\begin{document}
\draft

\title{Entropy production and fidelity for quantum many-body systems with noise}

\author{F.M.Izrailev and A.Casta\~neda-Mendoza}
\address{Instituto de F\'{\i}sica, Universidad Aut\'{o}noma de Puebla,
Apdo. Postal J-48, Puebla, Pue. 72570, M\'{e}xico}

  \date{\today}
  \maketitle

\begin{abstract}
We study dynamical properties of systems with many interacting
Fermi-particles under the influence of static imperfections. Main
attention is payed to the time dependence of the Shannon entropy
of wave packets, and to the fidelity of the dynamics. Our question
is how the entropy and fidelity are sensitive to the noise. In our
study, we use both random matrix models with two-body interaction
and dynamical models of a quantum computation. Numerical data are
compared with analytical predictions.
\end{abstract}

\section{INTRODUCTION}
\label{sect:intro}

There are two sources for the appearance of statistical properties
in quantum isolated systems. The first one which serves as a base
for the traditional statistical mechanics, is the thermodynamic
limit in which the number of particles is infinitely large. In
this case even integrable systems may be treated statistically.
The origin of this effect is an infinite number of independent
frequencies in the dynamics of a system. It is clear that in this
case any small interaction between particles, or with a heat bath,
gives rise to strong statistical properties.

Another situation occurs when the number of particles is finite
and not very large. In this case the role of interaction is
crucial, and chaos arises under some conditions. For systems with
the well defined classical limit, this chaos is known as {\it
quantum chaos} and can be compared with the classical one. This
situation is known in literature as {\it many-body chaos}, in
contrast to the thoroughly studied case of {\it one-body chaos}
which refers to a single particle in an external field. The well
known example is the hydrogen atom in a strong magnetic field.

In many-body systems the density of many-particle energy levels
increases extremely fast (typically, exponentially), both with an
increase of the number of particles and excitation energy. For
this reason, the interaction between particles can lead to a
strong mixing between many-particle basis states, thus resulting
in {\it chaotic eigenstates}. The latter term refers to the fact
that the components of such eigenstates can be practically treated
as pseudo-random variables, this leads to strong statistical
properties such as the relaxation of the system to a steady-state
distribution. Typical examples of such systems are compound
nuclei, complex atoms, atomic clusters, isolated quantum dots,
etc. Recent calculations for complex atoms \cite{FGGK94},
multicharged ions \cite{ions}, nuclei \cite{nuclei}, Bose-Einstein
condensates \cite{BBIS04}, and spin systems \cite{Nobel,spins}
have confirmed the dynamical origin of statistical laws in
isolated systems (see details in \cite{FI97,I00} and the review
\cite{kota}).

The onset of chaos for highly excited states and for many-particle
spectra has been thoroughly studied by making use of the Two-Body
Random Interaction (TBRI) model which was invented long ago
\cite{old}. In this model all {\it two-body} matrix elements are
assumed to be independent and random variables, however, the model
is essentially different from standard random matrix models where
the two-body nature of interaction is not taken into account. One
of the important results obtained in the frame of this model (see,
for example, \cite{GMW99} and references therein), is the
Anderson-like transition which occurs in the space determined by
many-particle states of the unperturbed part $H_0$. Due to a
random character of the perturbation, this delocalization
transition can be treated as the transition to a global chaos.
Above some critical value of the perturbation, the number $N_{pc}$
of principal components in eigenstates is typically large. For
example, $ N_{pc} \approx 150$ in the Ce atom \cite{FGGK94} and  $
N_{pc} \approx 10^4-10^5 $ in heavy nuclei \cite{nuclei}.

Recently, the theory of many-body chaos has been extended to
quantum computers. One can naturally expect \cite{dima0,dima} that
due to a very high density of energy levels, any kind of
perturbation may lead to decoherence effects thus destroying the
quantum computation. For this reason it is very important to
search for the conditions when the role of chaos can be
significantly reduced \cite{BBIT,DISS04}. So far, the study of the
many-body chaos has been mainly restricted by statistical
properties of the energy spectra and eigenstates. On the other
hand, in view of experimental applications, one needs to know what
are the {\it dynamical} properties of quantum systems with
strongly interacting particles. Especially, this is very important
for quantum computation \cite{F2000}, for which quantum protocols
assume a high stability of a long-time dynamics under the
influence of an environment or any kind of static imperfections.
Below, we analyze how the static disorder influences the dynamical
properties of quantum systems with complex behavior.

\section{Dynamics of systems with two-body interaction}

In what follows we assume that the models under consideration can
be represented by the Hamiltonian separated in two parts,
\begin{equation}
\label{H}H=H_0+V
\end{equation}
where $H_0$ describes the ``unperturbed" part and $V$ stands for
the ``perturbation". Then $V$ can be represented as
\begin{equation}
\label{V} V_{lk} = \langle l| V |k\rangle,
\end{equation}
where $\left| l\right\rangle $ and $\left| k\right\rangle $ stand
for {\it basis states} of $H_0$ (or, ``unperturbed" states).
Correspondingly, {\it exact states} $\left| \alpha \right\rangle $
of the total Hamiltonian $H$ are expressed as
\begin{equation}
\label{exact} |\alpha \rangle = C_k^\alpha |k\rangle.
\end{equation}
The coefficients $C_k^\alpha$ give the expansion of an exact state
in terms of the basis states (for $\alpha$ fixed), or the
expansion of a basis state in terms of the exact states (for $k$
fixed).

In the case when the Hamiltonian $H$ corresponds to a system of
interacting Fermi-particles, basis states can be constructed as
$\left| k\right\rangle =a_{k_1}^{+}...a_{k_n}^{+}\left|
0\right\rangle$ from the ground state $\left|0\right\rangle$ with
$a_s^{+}$ as the creation operator. In application to quantum
computer models the Hamiltonian $H_0$ typically describes a number
of non-interacting {\it qubits}, and $V$ stands for the
inter-qubit interaction needed for quantum computation (for other
variants, see below). In this case the basis state $|k\rangle$ is
a product of {\it single qubit states}, with $a_s^{+}$ as the
spin-raising operator (if the ground state $\left|0\right\rangle$
corresponds to all spins down).

In principle, the knowledge of the {\it state matrix} $C_n^\alpha$
and of the corresponding energy spectrum $E^\alpha$ gives a
complete information about the system. In particular, if an
initial state $\Psi(0)$ is some basis state $|k_0\rangle$, the
evolution of the $\Psi-$function is described by the expression,
\begin{equation}
\label{psit}\Psi (t) =\sum\limits_{n,m} C_n^\alpha
C_m^\alpha\left| \Psi(0)\right\rangle \exp(-iE^\alpha t).
\end{equation}
Here and below we assume that $\hbar=1$. As one can see, the
probability
\begin{equation}
\label{W0} w_m=|A_m|^2 =|\left\langle m|\Psi(t)\right\rangle|^2
\end{equation}
to find the system at time $t$ in the state $|m\rangle$ is
determined by the amplitude
\begin{equation}
\label{ampli} A_m= \left\langle m|\exp(-iHt)|m\right\rangle=
\sum\limits_\alpha|C_m^\alpha|^2\exp(-i E^\alpha t).
\end{equation}
When the perturbation is weak, only the component $A_m$ with
$m=k_0$ is large, therefore, the evolution of the system is close
to the periodic one. However, with an increase of the perturbation
strength, the number of components $A_m$ with relatively large
amplitudes increases, and the dynamics of wave packets can be
quite complicated. In order to characterize the dynamics of a
system in the unperturbed basis $|k\rangle$, one uses different
quantities. The main interest is to known how many unperturbed
states are involved in the dynamics depending on time. One of the
commonly used quantities is the Shannon entropy of the packet,
\begin{equation}
\label{S}S(t)=-\sum\limits_m w_m\ln w_m=-W_0\ln
W_0-\sum\limits_{m\neq 0}w_m\ln w_m \,;\,\,\,\,\,\,\,\,\sum_m w_m
=1.
\end{equation}
Here we introduced the {\it return probability} $W_0(t)=\left|
A_0(t)\right| ^2$ for the system to be in the initial state
$|k_0\rangle$, due its special role in global properties of the
dynamics. One can see that the effective number $N_p$ of
unperturbed states involved in the dynamics can be estimated as
$N_p \approx \ln S(t)$.

Another quantity which nowadays is under the close investigation,
is the so-called {\it fidelity},
\begin{equation}
\label{fidelity} {\cal F}(t) = |\langle
\Psi_p(t)|\Psi_u(t)\rangle|^2
\end{equation}
where $\Psi_u(t)$ stands for the $\Psi-$function of the
unperturbed system, and $\Psi_p(t)$ for that of the perturbed one.
In the case when perturbation is weak, this quantity serves as a
measure of the sensitivity of the system to a given perturbation
\cite{P84}. There are many results obtained for the fidelity, both
analytical and numerical ones, obtained for different systems
(see, for example, \cite{EWLC02} and references therein). The main
interest is whether the fidelity can serve as a good indicator of
the quantum chaos. Note, that the fidelity depends both on the
type of the perturbation, and on the form of an initial packet. In
what follows, we discuss this quantity in application to systems
of interacting Fermi-particles and to a model of quantum
computation.

By comparing the definition (\ref{fidelity}) with Eqs.(\ref{W0})
and (\ref{ampli}), one can see that in the case when the initial
state is a specific basis state $|k_0\rangle$, the fidelity is
nothing but the return probability \cite{FI01a},
\begin{equation}
\label{fid} {\cal F}(t) = W_0(t)=\left| A_0(t)\right| ^2;
\,\,\,\,\,\,\,\,
A_0(t)=\sum\limits_\alpha|C_{k_0}^\alpha|^2\exp(-i E^\alpha t)
\approx \int P_{k_0}(E)\exp(-i Et)dE.
\end{equation}
Here we replaced the summation by the integration that can be done
when the number of components $C_{k_0}^\alpha$ is large for a
fixed $k_0$. Generically, these components fluctuate strongly,
therefore, they can be treated as pseudo-random quantities. In
fact, this condition of a large number of pseudo-random components
in exact eigenstates can be used as the definition of chaos in
quantum systems. In this case, the time dependence of $W_0(t)$ is
entirely determined by the Fourier representation of
$P_{k_0}(E)=P(E,E_{k_0})$ where $E$ is the energy of exact
eigenstates and $E_{k_0}$ is the energy corresponding to the
unperturbed state $|k_0\rangle$ . This quantity is known in the
literature as the {\it strength function} (SF), or, as the {\it
local spectral density of states},
\begin{equation}
\label{strength}P(E,E_{k_0})\equiv
\overline{|C_m^{(\alpha)}|^2}\rho (E).
\end{equation}
Here $\rho (E)$ is the density of states of the total Hamiltonian
$H$, and the average is performed over a number of states with
energies close to $E$. In next Section we discuss main properties
of the fidelity (\ref{fid}) in application to close system of
interacting Fermi-particles \cite{FI01a}.

\section{Interacting Fermi-particles}

\subsection{Two-body random interaction model}

Let us assume that our model consists of finite number $N_p$ of
Fermi-particles that occupy $M$ {\it single-particle states}
$|s\rangle$ characterized by the corresponding energies
$\epsilon_s$. Then the unperturbed Hamiltonian $H_0$ describes
non-interacting particles, and the interaction between particles
is represented by $V$. In this case the basis many-body states
$|k\rangle$ can be constructed by the Slater determinant, $\left|
k\right\rangle =a_{s_1}^{\dagger
}\,.\,\,.\,\,.\,a_{s_{N_p}}^{\dagger }\left| 0\right\rangle$, and
the Hamiltonian takes the form
\begin{equation}
\label{H0}H_0=\sum_{s=1}^{M} \epsilon_s\,a_s^{\dagger }a_s.
\end{equation}
Correspondingly, the interaction term can be represented in this
basis as
\begin{equation}
\label{VV}V=\frac 12\sum V_{s_1s_2s_3s_4}\,a_{s_1}^{\dagger
}a_{s_2}^{\dagger }a_{s_3}a_{s_4}
\end{equation}
where $a_{s_j}^{\dagger }$ and  $ a_{s_j}$ are the
creation-annihilation operators.

The total Hamiltonian $H=H_0+V$ with (\ref{H0}) and (\ref{VV})
appears in many physical applications such as complex atoms,
nuclei, atomic clusters etc. In fact, the form of $H$ discussed
above is known as the {\it mean field approximation} for complex
quantum systems of interacting particles. In this description, the
unperturbed part $H_0$ represents the zero-order mean field for
the excited states with the ground state $E_1$, and the {\it
residual} two-body interaction is given by $V$ . Therefore, the
single-particle energies $\epsilon _s$ in such applications are,
in fact, renormalized quasi-particle energies (see details, for
example, in Ref.\cite{F96}). The considered model turns out to be
very useful for understanding of the onset of many-body chaos due
to a two-body interaction.

For the analytical analysis one assumes that all two-body matrix
elements $V_{s_1s_2s_3s_4}$ are random Gaussian numbers with the
zero mean and the variance $V_0^{\,2}=\left\langle
V_{s_1s_2s_3s_4}^{\,\,\,2}\right\rangle $. Although this
assumption of the randomness is quite strong, however, it was
found that many of the properties of this TBRI model are quite
generic and occur in different dynamical (without any random
parameters) systems. The results obtained in the frame of this
model can be also extended to interacting Bose-particles, see, for
example, Ref.\cite{bose}

Without the loss of generality one can assume that the
single-particle spectrum is non-degenerate, with the constant mean
level spacing $d_0=\,\langle\epsilon _{s+1}-\epsilon _s\rangle=1$.
Therefore, the model is defined by four parameters,
$M,\,N_p,\,d_0$ and $V_0$. Note that the number $N$ of many-body
states increases very strongly with an increase of number of
particles $N_p$ and number $M$ of single-particles states,
$N=\frac{M!}{N_p!(M-N_p)!}$. For this reason even for a relatively
small number of particles the size of the Hamiltonian matrix is
large, and exact eigenstates may consist of many unperturbed basis
states, thus providing us with a possibility to describe the
system statistically.

Due to a two-body nature of the interaction, the Hamiltonian
matrix has features that are different from those of the standard
random matrices that are typically used when studying chaos in
quantum systems. Specifically, the $H-$matrix is sparse, the
off-diagonal matrix elements of $V$ in the representation of
many-particle basis states are not completely independent, and the
global structure is a band-like (see details in
Refs.\cite{FGI96,I00}). This makes the analysis more complicated,
on the other side, such matrices give a better description of real
physical systems, in comparison with standard random matrix
ensembles.

One of the important results obtained recently in the frame of
this model is the Anderson-like transition which occurs in the
Fock space determined by many-particle states of $H_0$ (see, for
example, \cite{GMW99} and references therein). The critical value
$V_{cr}$ for this transition is determined by the density of
states $\rho_f=d_f^{-1}$ of those basis states which are {\it
directly coupled} by the two-body interaction. When the
interaction is very weak, $V_0\ll d_f$ , exact eigenstates are
very close to the unperturbed ones, consisting of a small number
of basis states. With an increase of the interaction the number
$N_{pc}$ of principal basis components  in exact eigenstates
increases, and for $V_0\geq d_f$ the eigenstates may be treated as
pseudo-random ones. In this case the number $N_{pc}$ can be
estimated as $N_{pc} \sim\Gamma/D$ where $ \Gamma$ is the
spreading width of the strength function (\ref{strength}). In such
{\it chaotic eigenstates} any external weak perturbation is
exponentially enhanced. The enhancement factor can be estimated as
$\sqrt{N_{pc}} \propto 1/\sqrt{D}$, see Ref.\cite{enhancement}
This huge enhancement have been observed in experiments when
studying the parity violation in compound nuclei (see, for
example, review \cite{W}). In what follows, we discuss the
dynamics of the model above this threshold of chaos.

\subsection{Fidelity}

We start now with the expression for the fidelity (\ref{fid}) at
small times. According to the perturbation theory, one can easily
get,
\begin{equation}
\label{smallT}W_0(t)\approx 1-\Delta_E^2t^2
\end{equation}
where $\Delta _E$ is the width of the strength function (SF) in
the energy space determined as
\begin{equation}
\label{deltadef}\Delta_E^2 = \sum_{m\neq k_0} V_{m,k_0}^2.
\end{equation}
The second moment $\Delta_E^2$ of the SF for the TBRI model can be
found explicitly \cite{FI97},
\begin{equation}
\label{SFwidth} \Delta_E^2=\frac{1}{4} V_0^2
N_p(N_p-1)(M-N_p)(M-N_p+3)
\end{equation}
where $V_0^2$ is the variance of the off-diagonal matrix elements
of the two-body interaction $V$. It is interesting that for
Fermi-particles the width $\Delta_E$ turns out to be independent
of a specific basis state $\left|k_0 \right \rangle$.

The expressions (\ref{smallT}) and (\ref{deltadef}) are universal,
they are correct for any kind of the perturbation $V$. Thus, the
quadratic time dependence of the fidelity on a small time scale
occurs independently on whether the dynamics is chaotic or
regular. This is understandable  since the difference between
periodic and chaotic motion can be detected on a large time scale
only (in principle, for $t \rightarrow \infty$).

The situation is much more complicated for large times. In order
to understand how strong the decrease of the fidelity on a large
time scale, one needs to know the form of the strength function in
the energy space. For a long time it was assumed that the SF has
generically the Breit-Wigner form (or, the same, the Lorentzian
form).  On the other hand, recently it was understood that in many
physical applications the form of the SF can be quite close to the
Gaussian form (see, e.g. \cite{FI01b}). This is because in
contrast to full random matrices, physical interactions $V$ always
have a finite width in the basis of $H_0$. This point was the
reason for Wigner to introduce the so-called Band Random Matrices
(BRM) with a finite width. Therefore, one more control parameter
arises which is the energy associated with the width of
perturbation, in addition to the width of the Lorentzian (see, for
example, Ref.\cite{FI01b} and references therein). Full analytical
treatment of the form of the SF for the BRM is given in
Ref.\cite{yan} Specifically, it was shown that with an increase of
the perturbation the form of the SF changes from the Lorentzian to
the Gaussian (apart from extremely strong perturbation, when the
non-physical semicircle form arises).

Similar transition from the Lorentzian to the Gaussian has been
analytically found for the TBRI model discussed above\cite{FI00},
although the analytical expression in the closed form is not
known. In order to evaluate the fidelity, in Ref.\cite{FI01a} the
following phenomenological expression has been suggested,
\begin{equation}
\label{approx} P(E)=B\frac{ \exp\left[-\frac{(E-E_0)^2}{2
\sigma^2}\right]}{(E-E_0)^2+\Gamma^2/4}
\end{equation}
where $E_0$ is the energy that corresponds to the basis state
$|k_0\rangle $. Strictly speaking, this formula is valid near the
center of the energy spectrum that has the Gaussian form,
otherwise one should take into account additional distortion
effects\cite{FI97}. Due to the normalization conditions, $\int
P(E) dE =1 $ and $\int E^2 P(E) dE = \Delta_E^2$, only one of
three parameters $B,\,\,\Gamma$ and $\sigma$ is free. The
relations between these parameters are:
\begin{equation}
\label{18}\frac 1B=2\left[ 1-\Phi \left( \frac \Gamma
{\sigma\sqrt{8} }\right)\right] \frac \pi \Gamma \exp
\left(\frac{\Gamma ^2}{8\sigma ^2} \right),
\end{equation}
and
\begin{equation}
\label{19}\Delta_E^2=B\left\{ \sigma\sqrt{2\pi } -\frac{\pi \Gamma
}2\exp \left( \frac{\Gamma ^2}{8\sigma ^2}\right) \left[ 1-\Phi
\left(\frac \Gamma {\sigma\sqrt{8}}\right)\right] \right\}
\end{equation}
where $\Phi (z)$ is the error function.

In the case of a relatively small (but non-perturbative)
interaction, the form of $P(E)$ is close to the Lorentzian. In
this case $\Gamma \ll \sigma $ and $\Gamma$ plays the role of the
half-width $\Gamma_0$ of the Lorentzian, $\Gamma \approx
\Gamma_0$. On the other hand, for a strong perturbation the SF has
the Gaussian form with $\sigma \approx \Delta_E$ and formally
$\Gamma \gg \sigma$. The value of $\Gamma_0$ for the TBRI model is
determined by the Fermi golden rule,
\begin{equation}
\label{GammaH}\Gamma _0(E)\simeq 2\pi \overline{\left|
V_{k_0f}\right| ^2} \rho_f(E)
\end{equation}
where $\rho_f(E)$ is the density of states directly coupled to the
basis state $|k_0\rangle$ by the two-body interaction $V$.

In the case of a relatively weak interaction (the Lorentzian form
of the SF) for large time, $t\gg t_c=1/\sigma $, the evaluation of
the fidelity gives,
\begin{equation}
\label{Lor} W_0(t)\approx\exp\left( \frac 1\pi \frac{\Gamma_0
^2}{\Delta_E^2}-\Gamma_0 t\right) \approx \exp\left(-\Gamma_0
t\right)
\end{equation}
where the correction term $\frac 1\pi \frac{\Gamma
^2}{\Delta_E^2}\approx \frac{2\Gamma }{\sigma \sqrt{2\pi }}$ is
small.

In the other limit case of a strong interaction (the Gaussian form
of the SF), $\Gamma \gg \sigma \approx \Delta_E$, we have $t_c\sim
\frac \Gamma {\sigma ^2} \gg \frac 1\sigma $. Therefore, the
leading dependence of $W_0(t)$ is the Gaussian,
\begin{equation}
\label{gau} W_0(t) \simeq \exp (-\Delta_E^2t^2),
\end{equation}
and only for a very long time $t\gg \frac \Gamma {\sigma ^2}$ it
becomes the exponential function \cite{FI01a},
\begin{equation}
\label{Wfinal}W_0(t)\approx \frac{\pi ^2\Gamma ^2}{8\Delta_E^2}
\exp \left( \frac{1}{4} \frac{\Gamma ^2}{\Delta_E^2}-\Gamma
t\right)
\end{equation}
It is important to note that in this case the return probability
$W_0(t)$ has large correction factor $\exp \left( \frac{1}{4}
\frac{\Gamma ^2}{ (\Delta E)^2}\right) $, in addition to the
standard decay law $W_0(t)=\exp (-\Gamma t)$.

\subsection{Shannon entropy}

For the Shannon entropy (\ref{S}) one needs to know the time
dependence of all amplitudes $w_m(t)$, see Eq.(\ref{W0}). In order
to find $w_m(t)$, in Ref.\cite{FI01b} the theory has been
developed which is based on the representation of the dynamics of
the system as a flow in the Fock space of  many-body states. This
approach can be compared with that developed in Ref. \cite{AGKL97}
where the dynamics in the many-particle basis was represented by a
flow on the Caley tree. For the exponential dependence
$W_0(t)=\exp(-\Gamma t)$, the solution for $W_n(t)$ has simple
form,
\begin{equation}
\label{sol}W_n(t)=\frac{ (\Gamma t)^n}{n!} W_0(t).
\end{equation}
As one can see, the dynamics of the system can be expressed in
terms of the return probability $W_0(t)$ which is a particular
case of the fidelity. Note, that the quantity $\Gamma$ stands here
for the effective width of the SF, and can be either the
half-width of the Lorentzian ($\Gamma=\Gamma_0$), or the square
root of the variance of the SF, ($\Gamma = \Delta_E$), depending
on its form (Lorentzian or Gaussian).

As a result, the expression for the Shannon entropy reads,
\begin{equation}
\label{s(t)} S(t)\approx  \Gamma t \ln N_f + \Gamma t -e^{-\Gamma
t} \sum\limits_{n=0}^\infty \frac {(\Gamma t)^n}{n!} \ln \frac
{(\Gamma t)^n}{n!},
\end{equation}
where $N_f$ is the number of basis states directly coupled to the
basis state $|k_0\rangle$. Two last terms in the right-hand-side
of Eq.(\ref{s(t)} ) turn out to be smaller than the first one,
therefore, one can write,
\begin{equation}
\label{Sff} S(t)\approx \Gamma t \ln N_f \{1 + f(t)\} \approx
\Gamma t \ln N_f
\end{equation}
with some function $f(t) \ll 1$ which slowly depends on time. In
this estimate for the increase of entropy, the influence of
fluctuations of $w_m$ is not taken into account. It can be shown
that for the gaussian fluctuations of the coefficients $A_m$ with
the variance given by their mean-square values, and $N_{pc}(t)
\approx \exp(S(t))\gg 1$, the entropy has to be corrected by a
small factor of the order of $\ln 2$ (see, for example,
\cite{I90}).

One should note that at small times the expression for the entropy
has to be modified since on this time scale the function $W_0(t)$
has always the form $W_0(t)=\exp (-\Delta_E^2\,t^2)$. To do this,
one needs to replace $\Gamma t$ in Eq.(\ref{sol}) by a more
accurate expression, $-\ln (W_0)$. This gives $\Delta_E^2\,t^2$
for small times, $t\ll \Gamma /\Delta_E^2$, and $ \Gamma t\,$ for
large times, $t\gg \Gamma /\Delta_E^2$. Therefore, at small times,
$t\ll \Gamma /\Delta_E^2\,$ the entropy is given by the expression
\cite{FI01b},
\begin{equation}
\label{Sfinal}S(t) \approx \Delta_E^2 t^2\,\left( 1+ {\cal O}(\ln
(t^2) \right)).
\end{equation}

As one can see, the ``exact'' expression (\ref{s(t)}) for the
entropy is quite complicated. Instead, in Ref.\cite{FI01b} it was
proposed to use the following simple expression which gives good
approximation for systems with a small number of particles,
\begin{equation}
\label{Sappr}S(t)=-W_0(t)\ln W_0(t)\,-(1-W_0(t))\ln \left(
\frac{(1-W_0(t))}{ N_{pc}^{(m)}}\right)
\end{equation}
Here $N_{pc}^{(m)}$ is the maximal value of the number of
principal components after the saturation of the entropy to its
maximal value. Quite often, this phenomenological parameter can be
estimated independently.

\section{The model of quantum computation}

\subsection{Description of the model}

Let us now apply the above analysis of the fidelity and entropy to
the model of quantum computation. This model describes a
one-dimensional chain of $L$ identical $1/2$-spins placed in an
external magnetic field. For a selective resonant excitation of
spins, it is suggested to choose the time independent part $B^z =
B^z(x)$ of a magnetic field  with the constant gradient along the
$x$-direction. This provides different Larmor frequencies for
different spins, $\omega_k = \gamma_n B^z=\omega_0+ak$, where
$\gamma_n$ is the spin gyromagnetic ratio and $ak=x_k$. One can
arrange relative directions of the chain with respect to the
$z$-axis in a way that the dipole-dipole interaction is
suppressed, and the main interaction between nuclear spins is due
to the Ising-like interaction mediated by the chemical bonds. This
model was first proposed in \cite{ber1} (see also
\cite{BDMT98,ber1.1,ber1.2}) as a simple realization of a
solid-state quantum computation.

In order to implement a time-dependent quantum protocol, the spins
are assumed to be subjected to a transversal circular polarized
magnetic field. Therefore, the total magnetic field has the
following form \cite{our,ber1.1,ber1.2},
\begin{equation}
{\vec B}(t)=[b^p_\perp\cos(\nu_p t+\varphi_p),
-b^p_\perp\sin(\nu_p t+\varphi_p), B^z(x)].
\end{equation}
In the above expression, $b^p_\perp$, $\nu_p$, and $\varphi_p$ are
the amplitudes, frequencies and phases of a circular polarized
magnetic field, respectively. The latter is given by a sum of
$p=1,...,P$ rectangular time-dependent pulses of length
$t_{p+1}-t_p$, rotating in the $(x,y)-$ plane, in order to make
selective excitations of specific spins. Thus, the quantum
Hamiltonian of the system has the form,
\begin{equation}
{\cal H}= -\sum\limits^{L-1}_{k=0} \left[\omega_kI^z_k+2
\sum\limits_{n > k}J_{k,n} I^z_k I^z_n \right]-
{{1}\over{2}}\sum\limits_{p=1}^{P}\Theta_p(t)\Omega_p
\sum\limits_{k=0}^{L-1} \Bigg(e^{-i\nu_p t-i\varphi_p}I^-_k+
e^{i\nu_p t+i\varphi_p}I^+_k\Bigg), \label{ham00}
\end{equation}
where the ``pulse function" $\Theta_p(t)$ equals $1$ during the
$p$-th pulse, for $t_p< t\le t_{p+1}$, otherwise, it is zero. The
quantities $J_{k,n}$ stand for the Ising interaction between two
qubits , $\omega_k$ are the frequencies of  spin precession in the
$B^z-$magnetic field, $\Omega_p$ is the Rabi frequency of the
$p$-th pulse, $I_k^{x,y,z} = (1/2) \sigma_k^{x,y,z}$ with $
\sigma_k^{x,y,z}$ as the Pauli matrices, and $I_k^{\pm}=I^x_k \pm
iI^y_k$.

For a specific $p$-th pulse, it is convenient to represent the
Hamiltonian (\ref{ham00}) in the coordinate system that rotates
with the frequency $\nu_p$. Therefore, for the time $t_p<t\le
t_{p+1}$ of the $p$-th pulse our model can be reduced to the {\it
stationary} Hamiltonian,
\begin{equation}
\begin{array}{ll}
{\cal H}^{(p)}=-\sum\limits_{k=0}^{L-1} (\xi_k I^z_k+ \alpha
I^x_k-\beta I^y_k)- 2  \sum\limits_{n>k}^{}J_{k,n}I^z_k I^z_n,
\label{ham}
\end{array}
\end{equation}
where $\xi_k=(\omega_k-\nu_p)$, $\alpha=\Omega_p \cos \varphi_p$,
and $\beta = \Omega_p \sin \varphi_p$.

Below we restrict our considerations by the Hamiltonian for a
single pulse, by choosing $\varphi_p=0$. We also assume a constant
interaction between nearest qubits, $J_{k,n}=J\delta_{k,k+1}$.
Then the Hamiltonian takes the form,
\begin{equation}
{\cal H}^{(p)}=\sum_{k=0}^{L-1} \Big [-\xi_k  I^z_k- 2J I^z_k
I^z_{k+1} \Big] -  \sum_{k=0}^{L-1} \Omega_p I^x_k .
\label{ham0}
\end{equation}

In the $z$-representation the Hamiltonian matrix of size $2^L$ is
diagonal for $\Omega_p=0$. Therefore, below we define ${\cal H}_0$
and $V$ as
\begin{equation}
\label{HV} {\cal H}_0=\sum_{k=0}^{L-1} \Big [-\xi_k  I^z_k- 2J
I^z_k I^z_{k+1} \Big] \,\,\,\,\,\,\,\,\,\,\,\,\,\, V= -
\sum_{k=0}^{L-1} \Omega I^x_k
\end{equation}
where we omitted the index $p$. The unperturbed basis (in which
${\cal H}_0$ is diagonal) is reordered according to an increase of
the index $s$ which is written in the binary representation,
$s=i_{L-1},i_{L-2},...,i_{0}$ (with $i_s=0$ or $1$, depending on
whether the single-particle state of the $i-$th qubit is the
ground state or the excited one). The parameter $\Omega$ thus is
responsible for a non-diagonal coupling, determining matrix
elements of $V$  as $V_{kn}=V_{nk}=- i\Omega/2$ with $n \neq k$.
As one can see, in contrast with the TBRI model discussed above,
in the $z-$ representation the interaction between particles is
absorbed by ${\cal H}_0$, and $V$ describes the coupling to the
external magnetic field.

There are two regions of the parameters of physical interest. The
first one is known as the so-called {\it non-selective} regime
which is defined by the conditions, $\Omega\gg \delta\omega_k \gg
J$. This inequality provides the simplest way to prepare a
homogeneous superposition of $2^L$ states needed for the
implementation of both Shor and Grover algorithms. The analytical
and numerical treatment of the model (\ref{HV}) in this regime has
shown\cite{our} that constant gradient magnetic field (with the
non-zero value of $a$) strongly reduces the effects of quantum
chaos. Namely, the chaos border turns out to be independent of the
number $L$ of qubits, in contrast to the models thoroughly studied
in Ref.\cite{dima} In particular, it was shown that quantum chaos
occurs for a very large coupling $\Omega = 100$ and very strong
interaction $J=100$ between qubits. For these parameters the
Wigner-Dyson distribution between nearest energy levels has been
observed, the fact that can serve as a numerical proof of the
quantum chaos. The transition to chaos for $\Omega = 100$ in
dependence on $J$ can be analytically understood with the use of
the transformation to the ``mean field basis" in which the
Hamiltonian is diagonal in the absence of the inter-qubit
interaction, $J=0$. In this basis the term with $J\neq 0 $ plays
the role of the interaction between $L$ blocks that correspond to
quantum numbers, and the structure of the Hamiltonian matrix is
similar to that of the TBRI model, see details in Ref\cite{our}.

Another region characterized by inequalities $\Omega_p \ll J_{k,n}
\ll a \ll\omega_k $, is referred as the {\it selective
excitation}, see Ref.\cite{ber1.1} In this regime each magnetic
pulse acts selectively on a chosen qubit, resulting in a resonant
transition. According to the quantum protocol, many such resonant
transitions take place for different $p$ pulses, with different
values of $\nu_p=\omega_k$. The detailed analytical analysis
\cite{select} has revealed that in this regime the perturbation
theory works very well for many pulses, thus, indicating that
there is no any effect of the quantum chaos. Therefore, the
implementation of the constant gradient magnetic field is very
effective in reducing any kind of decoherence.

\subsection{Shannon entropy and fidelity}

Now we discuss numerical data for the above model of quantum
computation, by paying the main attention to the time-dependence
$S(t)$ of the Shannon entropy and fidelity ${\cal F}(t)$. As we
have discussed, the time-dependence of the Shannon entropy
provides us with the information about the evolution of the wave
packets in the unperturbed basis. Since the latter is determined
by the Hamiltonian ${\cal H}_0$ in Eq.(\ref{HV}), the
time-dependence of the entropy is directly related to the
$\Omega-$dependent term. The main interest is in the role of the
interaction $J$, as well as of static random terms which we add to
$\Omega$. Specifically, each of non-zero off-diagonal elements is
now the sum of two terms, $\Omega \Rightarrow \Omega_0+\xi_p$
where $\xi_p$ stands for random variables with the zero mean and
variance $\sigma_p^2 =\langle \xi^2_p\rangle$. This kind of
randomness corresponds to the imperfections in the Rabi
frequencies.

   \begin{figure}
   \begin{center}
   \begin{tabular}{c}
   \includegraphics[height=10cm]{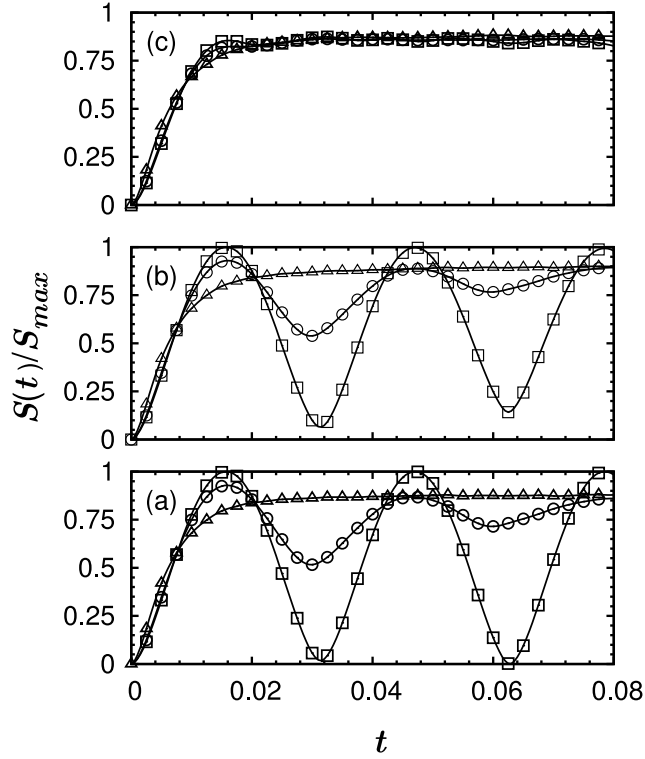}
   \end{tabular}
   \end{center}
   \caption[example]
   { \label{fig:fig1}Normalized Shannon entropy for $L=8, a=1, \Omega_0=100$, and
   different values of $J$: (a) $J=0$, (b) $J=10$, (c) $J=100$.
   Curves with squares correspond to the perturbation $\sigma_p=5$,
   with circles to $\sigma_p=10$, and with triangles to
   $\sigma_p=20$.}
   \end{figure}

In the first line, we discuss how the Shannon entropy depends on
time if the initial state is the basis state chosen, for
simplicity, at the center of the energy spectrum. Typical
dependencies for $S(t)$ are shown in Fig.1 where we have
normalized the entropy to its maximal value $S_{max}$ determined
by the total number of many-body states. The value $S_{max}$
corresponds to the case when all basis states are equally
populated, with the standard normalization that the total
probability is one. As one can see from the data, the entropy
saturates to its maximal value in two cases. The first one occurs
for the dynamical chaos which is due to a large value of the
inter-cubit interaction, $J=100$. In this case there is no
difference whether we have disorder $\sigma_p$ or not. Another
case is when the disorder is strong, $\sigma_p=20$. In this
situation, the saturation occurs even for $J=0$. These results are
quite instructive since they demonstrate equivalence of the
dynamical chaos to a disorder. It should be pointed out that the
entropy does not reach the maximal value $S(t)=1$ (in the
saturation) which may be explained by a suppressed chaos for the
states with the energies close to the edges of the spectrum. The
most important result is that for strong chaos (dynamical or due
to the disorder), the system can be well described by statistical
methods.

The dept $S_{min}$ of the first minimum in the time dependence of
$S(t)$ can be used as a relative measure of chaos in the system.
In Fig.2 the ratio $S_{min}/S_{max}$ is plotted against $J$, thus
showing the transition to chaos. Surprisingly, the transition
turns out to be quite smooth, in contrast to the transition
measured in terms of an effective number of components in exact
eigenstates\cite{our}. The latter transition has revealed two
borders, one is due to the ``delocalization" of the eigenstates in
the unperturbed basis (for $J \geq 15$), and the second one which
is due to the quantum chaos in the eigenstates (for $J \sim 100$,
when the level spacing distribution has the Wigner-Dyson form),
see details in Ref\cite{our}. Therefore, the Shannon entropy can
serve as the indicator of the delocalization, rather than of the
quantum chaos. Indeed, one can have a good relaxation even in a
completely integrable system, if one uses the basis which is
``very far" from the basis corresponding to the total Hamiltonian.

Let us now analyze how good is the correspondence of the above
data to the analytical expression (\ref{Sff}) obtained for the
TBRI model. This expression gives the linear increase of the
entropy before the saturation. The value of $\Gamma$ in
Eq.(\ref{Sff}) can be expressed via the variance (\ref{deltadef}),
since in our case $V$ is formally large ($\Omega =100$). Taking
into account that in each row of the matrix $V$ there are only
$N_f=L$ non-zero elements, one can easily compare the expression
(\ref{Sff}) with the numerical data, see Fig.3. One can see that
when the disorder is absent or small, $\sigma_p=0;\,15$, the
linear slope of $S(t)$ is slightly different from that predicted
by Eq.(\ref{Sff}). On the other hand, when the disorder is strong,
$\sigma_p=50$, the correspondence between the analytical
expression and numerical data is quite good. Thus, the theory
developed in the frame of the TBRI model appears to be also valid
for the models that are not fully random. Similar effect was also
observed in the Bose-Einstein condensate model \cite{bose}.
Namely, for a relatively strong interaction between bosons, the
Shannon entropy of the wave packet which was initially in the
condensate, was found to increase linearly with time in a good
agreement with the analytical expression (\ref{Sff}). This proves
an effectiveness of the approach developed in Ref.\cite{our} in
application to both dynamical and random systems.

   \begin{figure}
   \begin{center}
   \begin{tabular}{c}
   \includegraphics[height=7cm]{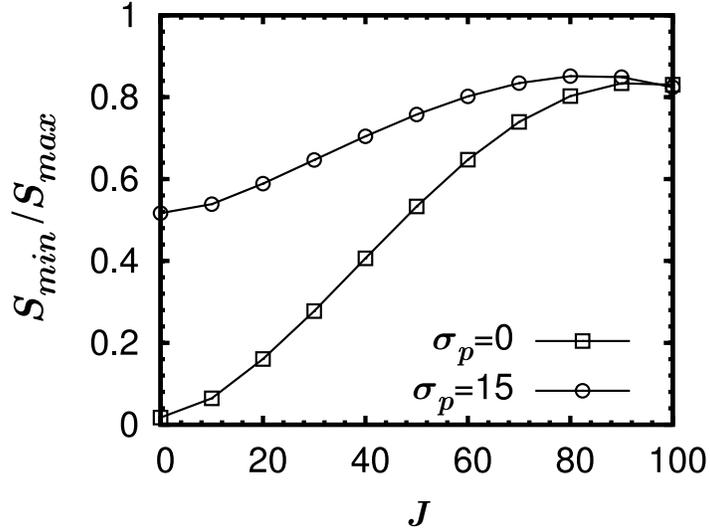}
   \end{tabular}
   \end{center}
   \caption[example]
   { \label{fig:fig2}
The ratio of the entropy in its first minimum, to the maximal
value $S_{max}$ in dependence on $J$ for two values $\sigma_p$ of
the disorder.}
   \end{figure}

   \begin{figure}
   \begin{center}
   \begin{tabular}{c}
   \includegraphics[height=7cm]{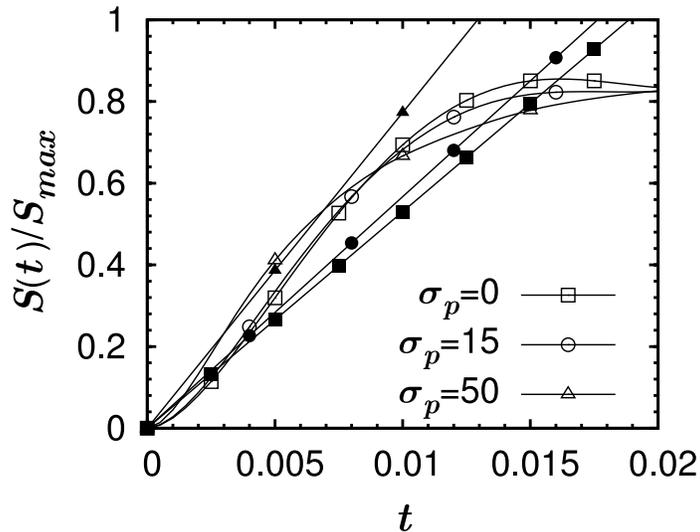}
   \end{tabular}
   \end{center}
   \caption[example]
   { \label{fig:fig3}
Small time scale for the Shannon entropy where the linear increase
is expected before the saturation. The data are given for
$\Omega_0=100,\,J=100,\,a=1$ and different strengths of the
disorder, $\sigma_p=0;\,15;\,50$. Open symbols stand for the
numerical data, full symbols, for the analytical estimate
(\ref{Sff}).}
   \end{figure}

We turn now to the fidelity. Since in our model there is a large
off-diagonal part determined by the constant term $\Omega$ and by
random terms $\sigma_p$, we introduce an additional random (small)
terms $\varepsilon_p$ that can be treated as the perturbation. In
this way we can analyze the dependence of the fidelity on time $t$
and on the perturbation $\varepsilon^2 = \langle
\varepsilon_p^2\rangle$ for different values of $J$ and
$\sigma_p$. Typical dependencies for the fidelity ${\cal F}(t)$
are reported in Fig.4. First, it is interesting to discuss the
data for the dynamical model, when $\sigma_p=0$. Unexpectedly, for
the integrable case ($J=0$) the fidelity is much less than for the
quantum chaos ($J=100$). This fact may have the same origin that
was discovered in Ref\cite{prosen}. However, we should stress that
our model has no classical limit, therefore, we can speak about
integrability or chaos without the reference to the classical
dynamics. Moreover, in Ref.\cite{prosen} the effect of a stronger
stability for chaotic situation occurs mainly on a small time
scale. In our case the effect is clearly seen on a large time
scale as well. One should not also forget that for $J=0$ the
unperturbed spectrum is highly degenerate, therefore, the observed
effect of a stronger stability for larger values of $J$ may be
related to the break of the degeneracy.

Another effect is that practically there is no difference for the
values $\sigma_p=15$ and $\sigma_p=50$. This means that for
$\sigma_p\geq 15$ the disorder is quite strong and does not depend
on its strength. Indeed, for such a disorder, the difference
between integrable ($J=0$) and chaotic ($J=100$) cases is not big.
Thus, the results given by the fidelity confirm the conclusion
drawn from the entropy data: dynamical chaos that is due to a
strong interaction $J=100$ between qubits, and chaos due to a
strong disorder $\sigma=50$ are similar (compare fidelity for
$J=100, \sigma_p=0$ with that for $J=0, \sigma_p=50$).

   \begin{figure}
   \begin{center}
   \begin{tabular}{c}
   \includegraphics[height=10cm]{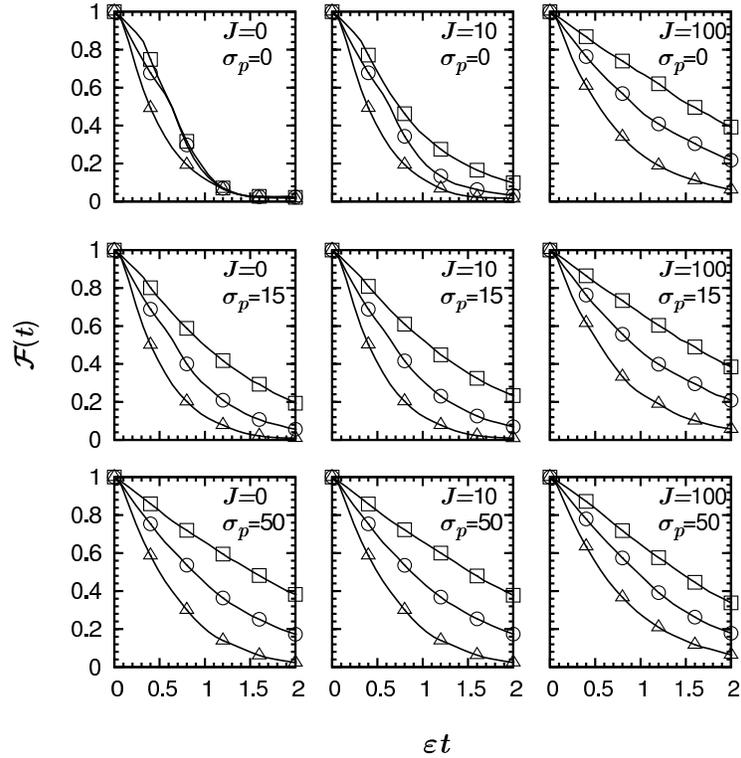}
   \end{tabular}
   \end{center}
   \caption[example]
   { \label{fig:fig4}
Fidelity for different inter-qubit interactions $J$ and the
disorder $\sigma_p$, with $\Omega_0=100$. The time dependence is
shown for 3 different values of the perturbation, $\varepsilon=5$
(squares), $\varepsilon=10$ (circles), and $\varepsilon=20$
(triangles).}
   \end{figure}

It is also instructive to analyze more carefully the
time-dependence of the fidelity for the dynamical case when
$\sigma_p=0$, see Fig.5. One can see a very good scaling ${\cal
F}={\cal F}(\varepsilon t)$ for the integrable case $J=0$, in
contrast with the chaotic case $J=100$. Moreover, as is seen from
the data in semi-log scale, practically for all values of ${\cal
F}$, apart from very small values, the time dependence is the
Gaussian , ${\cal F} \sim \exp(-C\varepsilon^2t^2)$. This is in
contrast to the chaotic case with $J=100$ (additional data show
that the dependence of ${\cal F}$ on time is closer to the
exponential one rather to the Gaussian one). The origin of a good
scaling, together with its Gaussian form is not clear, however,
one can expect that it is related to the form of the strength
function, as in the case of the TBRI model. To clear up this
problem, additional numerical study is needed.

   \begin{figure}
   \begin{center}
   \begin{tabular}{c}
   \includegraphics[height=10cm]{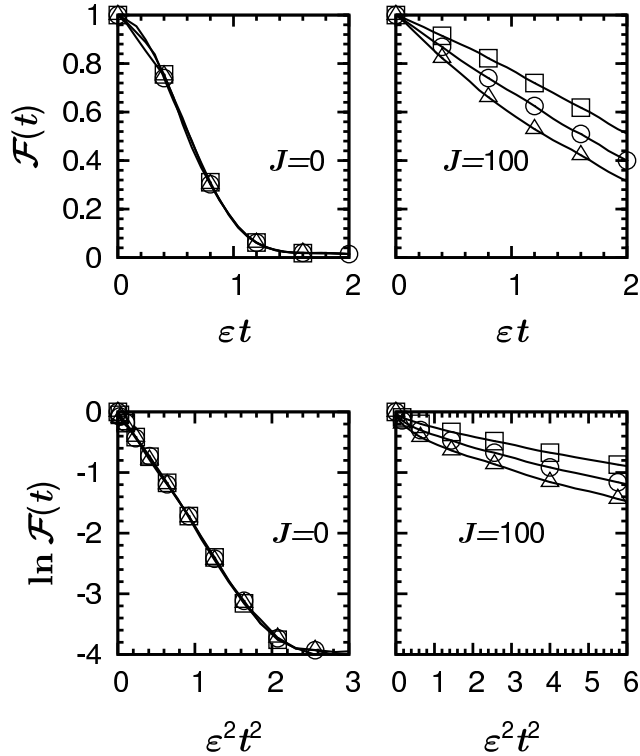}
   \end{tabular}
   \end{center}
   \caption[example]
   { \label{fig:fig5}
Comparison of the fidelity for the regular, $J=0$, and chaotic,
$J=100$, cases, with $\Omega_0=100$. In both cases the disorder is
absent, $\sigma_p=0$. Different scales are used for the fidelity
and $\varepsilon t$, in order to reveal the time-dependence on a
large time scale. As in Fig.4, $\varepsilon=3;\,5;\,7$ with
squares, circles and triangles, respectively.}
   \end{figure}

\section{Concluding remarks}

As is mentioned above, there are many studies where the
time-dependence of the fidelity is discussed from the viewpoint of
quantum chaos. The main attention was paid to a long time scale on
which an exponential or Gaussian decay typically occurs. In the
case of the exponential decay the question of general interest is
about the characteristic parameter of this decay. Specifically, it
is widely discussed whether the exponential decrease of the
fidelity is governed by the Fermi golden rule (in this case the
strength function has the Lorentzian form), or by the classical
Lyapunov exponent of the corresponding classical system.

On the other hand, in application to the models of quantum
computation the problem of a long-time behavior of the fidelity
seems to be irrelevant. Indeed, for a quantum computation one
needs to have a very stable regime where the fidelity is close to
one. Therefore, in this application one should pay the main
attention to the short time scale, where the fidelity differs from
1, say, not more than $10^{-3}$. This is because the time of the
quantum computation can be very large, and many pulses of an
external magnetic field are needed to implement the quantum
protocol. One should expect that in this situation the
perturbation theory works well (see Ref.\cite{select}). Let us see
what the perturbation theory gives for our model (\ref{HV})
describing the system within a single pulse. In our case the total
Hamiltonian $H=H_0+V$ has both diagonal and off-diagonal parts,
and let us assume that the perturbation is described by an
additional term $\Sigma$ of the same structure as the $\Omega-$
term in Eq.(\ref{HV}), with the variance $\varepsilon^2$ for its
matrix elements. Also, we assume that the interaction part $V$ may
have additional terms, see above.

In this case the perturbation theory gives,
\begin{equation}
\label{fid_new} {\cal F}(t) = 1- \delta_E^2 t^2 - Re \langle R
\rangle t^2
\end{equation}
where
\begin{equation}
\label{R} \delta_E^2 = \sum_{m\neq k_0} \Sigma_{m, k_0}^2=
\varepsilon^2 N_\varepsilon \,\,\,\,;\,\,\,\,\,\,\, R=H\Sigma -
\Sigma H.
\end{equation}
with $N_\varepsilon$ as the number of non-zero matrix elements of
$\Sigma$ for a fixed $k_0$.

One can see that when the unperturbed Hamiltonian $H$ has a
diagonal form (for example, when $V=0$) the terms $H$ and $\Sigma$
commute (therefore, $R=0$) and we come to the previous expression
(\ref{smallT}) with $\Delta_E^2 = \delta_E^2$. Since in our
numerical study the perturbation has the same structure as the
off-diagonal $\Omega-$ terms, one can get $R=V\Sigma - \Sigma V=0$
(assuming $\langle \Sigma_{m, k_0}\rangle=0$). Therefore, for our
particular type of the perturbation, there is no influence of the
term $V$. Our numerical data confirm these findings. The
expression (\ref{fid_new}), however, shows that in a general case
the fidelity strongly depends on the type of the perturbation.

In conclusion, we have studied the time dependence of the Shannon
entropy and fidelity for the model of a quantum computation, in
comparison with analytical predictions obtained for the model with
two-body random interaction. In spite of a big difference between
these two models (one is the dynamical one and another is random),
we have found that in many aspects some properties of the dynamics
are quite similar. One of the important results is that global
properties of the dynamics look the same both for the dynamical
model with strong chaos, and for the model with a strong disorder.
In particular, the entropy and fidelity behave in the same way,
manifesting the relaxation of the system to a statistical
equilibrium. The time scale on which this relaxation occurs can be
described by a linear increase of the entropy, with a good
correspondence with the simple analytical expression. This fact
confirms the expectation that the TBRI model can serve as the base
in understanding the properties of quantum many-body chaos. Our
numerical data for the fidelity do not confirm the expectation
that this quantity may serve as a good indicator of the quantum
chaos. Specifically, there are many open questions related to the
problem of the universal properties of the fidelity. Much depends
on the type of the perturbation and on the form of an initial
packet, therefore, more extensive studieds are needed. As for the
application of the fidelity to a quantum computation, one can
expect that real interest is restricted by a small time scale
where the standard perturbation theory works very well, and there
is no influence of the quantum chaos.


\end{document}